\documentclass[aps,prb, twocolumn,nofootinbib]{revtex4-1}

\usepackage{amsmath}
\usepackage{amsfonts}
\usepackage{amssymb}
\usepackage{dcolumn}
\usepackage{bm}
\usepackage{graphicx}
\usepackage{adjustbox}
\usepackage{epstopdf}
\usepackage{color}
\usepackage{xcolor}
\usepackage{textcomp}
\usepackage{gensymb}
\usepackage{adjustbox}
\usepackage{ulem}
\usepackage{comment}
\definecolor{deep-blue}{rgb}{0.17, 0.17, 0.89}

\def\BaOs  {Ba$_2$NaOsO$_6$}

    \def \etal {{\it et al.}}

\begin{document}
\title{Evidence for orbital ordering in Ba$_2$NaOsO$_6$, a Mott insulator with strong spin orbit coupling, from First Principles}
\author{R. Cong$^{1}$, Ravindra Nanguneri$^{2}$, Brenda Rubenstein$^{2, \dag}$, and V. F. Mitrovi\'c$^{1, \dag}$}
\address{$^{1}$ Department of Physics, Brown University, Providence, Rhode Island 02912, USA}
\address{$^{2}$ Department of Chemistry, Brown University, Providence, Rhode Island 02912, USA}

\date{\today}
\begin{abstract}
We present first principles calculations of the  magnetic  and orbital properties of Ba$_2$NaOsO$_6$ (BNOO), a 5$d^1$ Mott insulator with strong spin orbit coupling (SOC) in its low temperature emergent quantum phases. 
Our computational method takes into direct consideration recent  NMR results that established that BNOO develops a local octahedral distortion  preceding  the formation of long range magnetic order. We found that the two-sublattice canted ferromagnetic ground state identified in Lu \etal, Nature Comm.  {\bf 8}, 14407 (2017) is accompanied by  a two-sublattice staggered orbital ordering pattern in which the $t_{2g}$ orbitals are selectively occupied as a result of strong spin orbit coupling. The staggered orbital order found here using first principles calculations asserts the previous proposal of Chen \etal, Phys. Rev. B {\bf 82}, 174440 (2010) and Lu  \etal, Nature Comm.   {\bf 8}, 14407 (2017), 
that two-sublattice magnetic structure is the  very manifestation of staggered quadrupolar order. Therefore,  our results affirm the essential role of multipolar spin interactions in the  microscopic description of  magnetism in systems with locally entangled  spin and  orbital  degrees of freedom. 
 \end{abstract}

\maketitle

\section{Introduction}
\label{sec-introduction}
  \vspace*{-0.20cm}
  %
The competition between electron correlation and spin orbit coupling (SOC) present in materials containing 4- and 5$d$ transition metals is an especially fruitful tension predicted to lead to the emergence of a plethora of exotic quantum phases, including quantum spin liquids, Weyl semimetals, Axion insulators, and phases with exotic magnetic orders \cite{Kim08Nov, Chen_PRB_2010, Chen:2011, Ishizuka:2014el, trivedi_2017, balents_SOC_review_2014, Balents_2017,  J_eff_half_MI_2,Jeff_half_iridates,5d_IMT_CMS_RP_series,weyl, balents_SOC_review_2014}. 
There has been an active quest to develop microscopic theoretical models to describe such systems with comparably strong correlations and SOC to enable the prediction of their emergent quantum properties \cite{Kim08Nov, Chen_PRB_2010, Chen:2011, Ishizuka:2014el,   trivedi_2017}.  
In strong Mott insulators, mean field theories predict   strong SOC to partially lift the degeneracy of total angular momentum eigenstates by entangling orbital and spin degrees of freedom to produce highly nontrivial anisotropic exchange interactions \cite{Chen_PRB_2010, Chen:2011,Balents_2017, trivedi_2017}. These unusual interactions are anticipated to promote quantum fluctuations that generate such novel quantum phases as an unconventional antiferromagnet with dominant magnetic octuple and quadrupole moments and a noncollinear ferromagnet whose magnetization points along the [110] axis and possesses a two-sublattice structure. 

Because their SOC and electron correlations are of comparable magnitude\cite{balents_SOC_review_2014}, 5$d$ double perovskites (A$_{2}$BB$^{'}$O$_{6}$) are ideal materials for testing these predictions. Indeed,  recent  NMR measurements on a representative material of this class, Ba$_2$NaOsO$_6$ (BNOO), revealed  that it possesses a form of exotic ferromagnetic order: a two-sublattice canted ferromagnetic (cFM) state, reminiscent of theoretical predictions  \cite{Lu_NatureComm_2017}.  Specifically, upon lowering its temperature, BNOO evolves from a paramagnetic (PM) state with perfect $fcc$ cubic symmetry into a broken local symmetry (BLPS) state. This BLPS phase precedes the formation of long-range magnetic order, which at sufficiently low temperatures, coexists with the two-sublattice cFM order, with a net magnetic moment of $\approx 0.2 \, \mu_{B}$ per osmium atom along the [110] direction. One key  question  that  remains is whether such  cFM  order implies the existence of complex orbital/quadrupolar order. 

In this paper, we report a two-sublattice orbital ordering pattern that coexists with cFM order  in BNOO, as revealed by DFT+U calculations. Evidence for this order  is  apparent  in BNOO's selective occupancy of the $t_{2g}$ orbitals and spin density distribution. More specifically, the staggered orbital pattern is manifest in BNOO's partial density of states and band structure,  which possesses a distinct  $t_{2g}$ orbital contribution along high symmetry lines. This staggered orbital pattern is not found in  the FM[110] phase. The results of our first  principles  calculations  paint  a  coherent  picture  of the coexistence of cFM order with staggered orbital ordering in the  ground state of BNOO.  Therefore, the  staggered orbital order discovered here  validates  the previous proposal  that  the two-sublattice magnetic structure, which defines the cFM order in BNOO, is the  very manifestation of staggered quadrupolar order with distinct orbital polarization on  the  two-sublattices \cite{Chen_PRB_2010, Lu_NatureComm_2017}. Furthermore, our results affirm that   multipolar spin interactions are  an  essential ingredient of quantum theories of magnetism in  SOC materials.  

This paper is organized as follows. The details of  our  first principles (DFT) simulations  calculations of NMR observables is first described in Section~\ref{CompApp}.  In Section~\ref{OOrder}, we present our numerical results for the nature of the orbital order for a given imposed magnetic state.   Lastly, in Section~\ref{Conc}, we conclude with a summary of our current findings and their bearing on the physics of related materials.

\section{Computational Approach}
\label{CompApp}
  \vspace*{-0.20cm}

%
%
 All of the following computations were performed  using the Vienna Ab initio Simulation Package (VASP),  complex version 5.4.1/.4, plane-wave basis DFT code \cite{vasp_1,vasp_2,vasp_3,vasp_4} with the Generalized-Gradient Approximation (GGA) PW91 \cite{GGA}  functional and two-component spin orbit coupling. We used $500$ eV as the plane wave basis cutoff energy and we sampled the Brillouin zone using a 
$10\times 10\times 5$ $k$-point grid. The criterion for stopping the DFT self-consistency cycle was a $10^{-5}$~eV difference between successive total energies.   Two  tunable parameters, $U$ and $J$,  were  employed. $U$ describes the screened-Coulomb density-density interaction acting on the Os 5$d$ orbitals and $J$ is the Hund's interaction that favors maximizing $S^z_{total}$ \cite{hund}. In this work, we set $U=3.3$ eV and $J=0.5$ eV based upon measurements from Ref. 12 and calculations in Ref. 21. We note that these parameters are similar in magnitude to those of the SOC contributions we observe in the simulations presented below, which are between 1-2 eV. This is in line with previous assertions that the SOC and Coulomb interactions in 5$d$ perovskites are similar in magnitude. Projector augmented wave (PAW) \cite{PAW_Blochl,PAW_vasp} pseudopotentials (PPs) that include the $p$ semi-core orbitals of the Os atom, which are essential for obtaining the observed electric field gradient (EFG) parameters \cite{CongEFG19}, were employed to increase the computational efficiency. 
A monoclinic unit cell with P2 symmetry is required  to realize cFM order  (see the Supplemental Information). The lattice structure with BLPS characterized by the  orthorhombic  Q2 distortion mode  that was  identified  as being in the best agreement  with  NMR findings  and referred to  as  Model A.3
in \mbox{Ref. \cite{CongEFG19} }was imposed. 

   \begin{table}[b]
    \centering
    \begin{adjustbox}{max width=\columnwidth}
    \begin{tabular}{|c|c|c|c|c|c|c|c|c|}
    \hline
{ } & \boldmath{$|\vec{S}|$} & {\boldmath $\phi(S)$} & {\boldmath $|\vec{L}|$} & {\boldmath $\phi(L)$} & {\boldmath $M$} & {\boldmath $\phi(M)$} \\ \hline \hline
{\bf cFM} &{} & & & & &  \\ \hline
    {\bf  Os1} & 0.55 & -41.56 & 0.44 & 90+46.29 & 0.12 & -34 \\ \hline
    {\bf  Os2} & 0.55 & 90+28.73 & 0.43 & -31.07 & 0.11 & 110 \\ \hline
    {\bf FM110}    &{} & & & & &  \\  \hline 
    {\bf  Os} & 0.83 & 45 & 0.52 & 225 & 0.31 & 45 \\ \hline
    \end{tabular}
    \end{adjustbox}
    \caption{cFM and FM[110] magnetic moments for the imposed representative BLPS structure using GGA+SOC+U. The angles, $\phi$, are in degrees and measured anti-clockwise with respect to the +$x$ axis. The magnitudes of spin, orbital, and total moments are denoted by $|\vec{S}|$, $|\vec{L}|$, and $|\vec{M}|$, in units of $\mu_B$, respectively. The small net magnetic moment is due to the anti-aligment of $\vec{S}$ and $\vec{L}_{\rm eff}$ in the $J_{\rm eff}=\frac{3}{2}$ state. As of now,  the  
    FM110 state  has not been  experimentally identified in  a $5d$ double perovskite.  
          } 
 \vspace*{-0.4cm}
    \label{tab:cFM_FM110_magnetic_order_GGA_SOC_U}
\end{table}

The general outline  of  the calculations we performed is described in the following. 
 We first carried out single self-consistent or `static' calculations with GGA+SOC+U with a fixed BLPS structure for  Model A, representing the orthorhombic  Q2 distortion mode.     
Then, a magnetic order with [110] easy axes, as dictated by experimental findings \cite{erickson2007FM, Lu_NatureComm_2017}, is imposed on the osmium lattices. 
Typically, we  found that the final moments converged  to nearly  the same directions as the initial ones. 
Specifically, two types of such initial order are considered:  (a) simple FM order with   spins pointing along the [110] direction;  and, (b)
 non-collinear, cFM oder in which  initial magnetic moments are imposed  on the two osmium sublattices in the directions determined in 
 \mbox{Ref. \onlinecite{Lu_NatureComm_2017}}. We used the Methfessel-Paxton (MP) smearing technique \cite{MP_smearing} to facilitate charge density convergence.  For the density of states and band structure calculations, we employed the  tetrahedron smearing  with  Bl\"ochl corrections \cite{Blochl_tetrahedron_corrections} and  Gaussian methods, respectively.
 
 %
\begin{figure}[t]
  \vspace*{-0.0cm}
\begin{minipage}{0.98\hsize}
 \centerline{\includegraphics[scale=0.37]{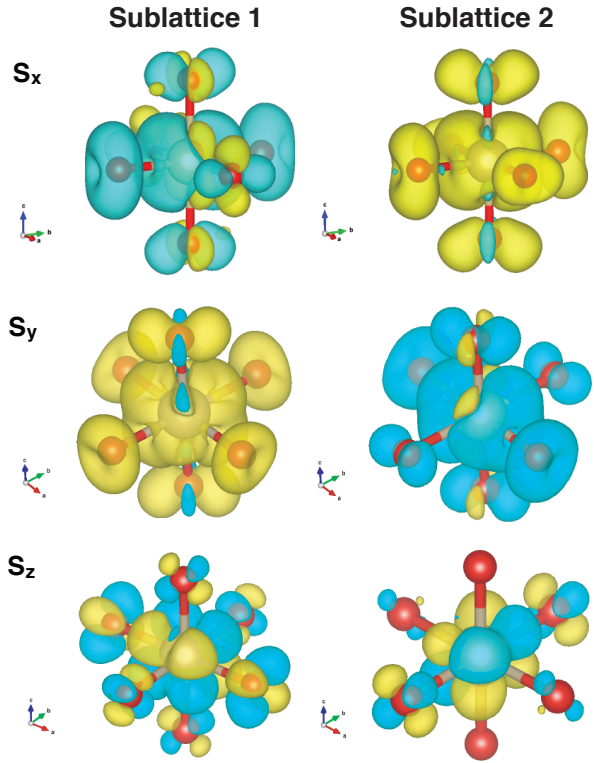}} 
\begin{minipage}{.98\hsize}
 \vspace*{-0.2cm}
\caption[]{\label{SpinDen} \small 
Contour plots of the spin density  on two distinct sublattices of the  BLPS structure (Model A.3 in  Ref.  \cite{CongEFG19})  from  GGA+SOC+U calculations. The S$_x$ (top row), $S_y$ (center row), and $S_z$ (bottom row) components of the spin density on a single Os octahedron from  sublattice 1 (left column) and sublattice 2 (right column) are plotted.
 The different colors denote the signs of the $S_{x,y,z}$ projections. The isovalues are blue for positive $S_{x,y,z}$, $0.001$, and yellow for negative $S_{x,y,z}$, $-0.001$. On the top left, the negative $S_x$ density is sandwiched between the lobes of the positive $S_x$ densities on the Os atom, and vice-versa for the Os atom on the top right. On the top left, four of the O atoms have a cloverleaf  spin density  pattern with alternating positive and negative $S_x$ densities, while on the top right, only the two axial O atoms have this pattern. The other O atoms in the top two OsO$_6$ octahedra have spin densities that are uniformly polarized. 
 }
 \vspace*{-0.6cm}
\end{minipage}
\end{minipage}
\end{figure}
%
  \vspace*{-0.10cm}

 \section{Orbital Ordering with  Imposed Magnetic cFM and FM110 Orders}
\label{OOrder}
  \vspace*{-0.20cm}

%
In the following subsections, we report our results for the orbital order, band structure, and density of states of BNOO when we impose  magnetic order with [110] easy axes and 
 the local orthorhombic  distortion  
 that best matched experiments \cite{CongEFG19}. 
In Table \ref{tab:cFM_FM110_magnetic_order_GGA_SOC_U}, we summarize the converged orbital and spin magnetic moments. In BNOO, \mbox{$M =2 S +L_{\rm eff} = 0$}, since   the t$_{2g}$ level can be regarded as a pseudospin with $L_{\rm eff} = -1$. The magnitude of the spin moment, $|\vec{S}|$, is in the vicinity of $\approx 0.5  \mu_B$, while the orbital moment, $|\vec{L}|$, is $\approx 0.4 \mu_B$. These values are reduced from their purely local moment limit due to hybridization with neighboring atoms, and, in the case of $\vec{L}$, by quenching caused by the distorted  crystal field.  For imposed cFM order, we find that the relative angle $\phi$ within the two  sublattices  is in agreement with our NMR findings in Ref. 12. Indeed, first principles  calculations,  performed outside of our group, taking into  account multipolar spin interactions  found that the reported canted angle of \mbox{$\approx 67 ^{{\circ}}$} corresponds to the global energy minimum \cite{thesisCF}.

%
Next, we will explore the nature of the orbital ordering.
Previous first principles works hinted at the presence of orbital order in BNOO, but did not fully elucidate its nature \cite{Pickett_2015}.
Since we imposed cFM order and SOC, we were able to obtain a more exotic orbital order than  uniform  ferro-order. We report below evidence for a type of layered, anti-ferro-orbital-order (AFOO) that has been shown to arise in the  mean field  treatment  of multipolar Heisenberg models with SOC \cite{trivedi_2017}. 

  First, we analyze the nature of the orbital order by computing the spin density. The spin density is a continuous vector field of the electronic spin, and can point in non-collinear directions. Its operators are the product of the electrons' density and their spin-projection operators, such as 
$\Delta^z(\vec{r}) = \sum_{i} \delta(\vec{r}_i - \vec{r}) S^z_i.$ 
%

 %
\begin{figure}[t]
  \vspace*{-0.0cm}
\begin{minipage}{0.98\hsize}
 \centerline{\includegraphics[scale=0.97]{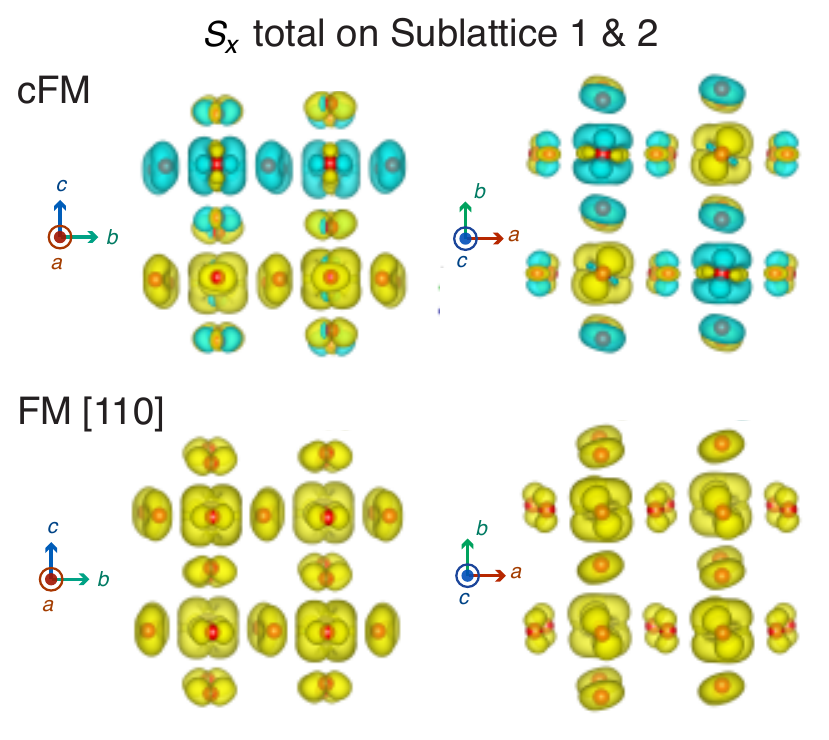}} 
\begin{minipage}{.98\hsize}
 \vspace*{-0.3cm}
\caption[]{\label{spinDensTot} \small 
Two views of the $S_x$-component of the spin density for imposed FM order and  an orthorhombic Q2 distortion (Model A.3 in Ref. 21 on both sublattices as viewed along the -$a$ and -$c$ directions. 
This component shows only the $S_x$-projection of the spin vector field. The isovalues are blue, positive S$_x$: $0.001$, and yellow, negative S$_x$: -$0.001$.  }
 \vspace*{-0.5cm}
\end{minipage}
\end{minipage}
\end{figure}
%

The spin densities are given by the expectation value, 
\vspace*{-0.2cm}
 \begin{equation}
 \vspace*{-0.2cm}
 \langle \Delta^z(\vec{r}) \rangle=Tr[\rho_d \Delta^z(\vec{r})] \, , 
 \end{equation}
 where $\rho_d$ is the $5d$-shell single-particle density matrix obtained from DFT+U calculations. 
 In  \mbox{Fig. \ref{SpinDen}},   
 the $\langle \Delta^{x,y,z}(\vec{r}) \rangle$,  obtained via GGA+SOC+U calculations, are displayed   for two distinct  Os sublattices.  
 The results  illustrate  that the spins are indeed localized about the Os atoms, and that there is a noticeable imbalance in the distribution of the $n_{\uparrow}$ and $n_{\downarrow}$ spin densities, which manifests in their difference, 
$\langle \Delta^{z}(\vec{r}) \rangle  \equiv \, n_{\uparrow}(\vec{r})-n_{\downarrow}(\vec{r})$.
 The difference in the spatial distribution between the two sublattice spin densities is indicative of the orbital ordering. 
 The net spin moments are obtained by integrating the spin density over the volume of a sphere enclosing the Os atoms. 
  
   In Fig.~\ref{SpinDen}, it is visually clear that: {\bf I.} The $S^x$ (top) and $S^y$ (center) spin density components are overwhelmingly of a single sign, which gives rise to net moments in the $(a,b)$ plane; and {\bf II.} The signs of $S^x$ and $S^y$ between the two sublattices are reversed, indicating that the sublattice spins are canted symmetrically about the [110] direction and the angle between them exceeds $90^{\circ}$.  In contrast, for $S^z$ (bottom), both signs of $S^z$ contribute equally, so that the net $S^z\approx 0$ after integrating over the sphere.

 \begin{center}	
 %
\begin{figure}[b]
  \vspace*{-0.0cm}
\begin{minipage}{0.98\hsize}
 \centerline{\includegraphics[scale=0.56]{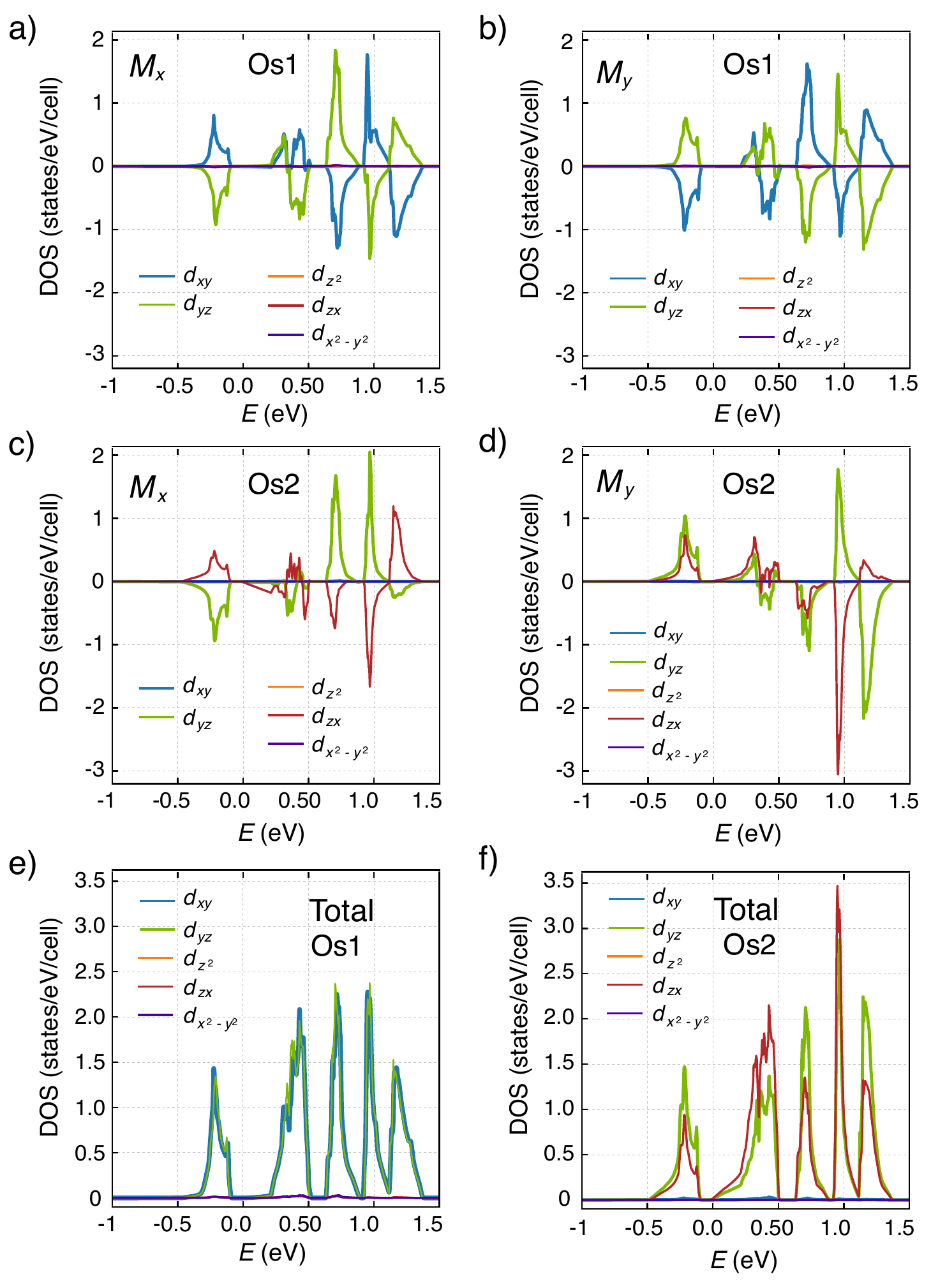}} 
\begin{minipage}{.98\hsize}
 \vspace*{-0.2cm}
\caption[]{\label{fig:pdos} \small 
The partial density of states (PDOS) for spin decomposed parts of the  Q2 orthorhombic distortion (Model A.3 in Ref. 21) for the Os atom in each sublattice, Os1 and  Os2.   
}
  \vspace*{-0.5cm}
\end{minipage}
\end{minipage}
\end{figure}
%
\end{center}
  \vspace*{-0.39cm}

In \mbox{Fig. \ref{spinDensTot}}, we plot the total $S_x$-component of the spin density over two sublattices for both types of imposed 
magnetic order. It is evident that the staggered orbital pattern only arises when cFM order is imposed. Therefore, we demonstrate that the staggered orbital order can solely coexist with cFM order.


We note also that in Figs.~\ref{SpinDen} and \ref{spinDensTot} there is non-negligible spin density on the O atoms of the OsO$_6$ octahedra. It is usually thought that atoms with closed shells, like O in stoichiometric compounds, possess negligible spin densities.
This is an unexpected feature in BNOO that has been previously noted in Ref.~\onlinecite{Pickett_2015}, and is due to the stronger $5d$-$2p$ hybridization, which results in OsO$_6$ cluster orbitals. The spin imbalance is a quantity associated with the cluster rather than the individual atoms, which is why we see the spin densities on the O atoms.

For non-collinear systems, the orbital character of each osmium's $5d$ manifold can be further decomposed into the Cartesian components of the spin magnetization: $\langle S_i \rangle \equiv   M_i$, $i=x,y,z$. Since the spins lie in the $(xy)$ plane and the  $M_{z}$ component is zero for both sublattices, we only  plotted $M_{x}$, $M_{y}$, and the total PDOS for the two sublattices. We see in \mbox{Fig. \ref{fig:pdos}} that, firstly, for both sublattices, only  the $t_{2g}$ orbitals  have an appreciable density of states  consistent with the fact that the calculated $d$  occupation  at  the  Os sites is $\langle n_d \rangle < 6$. Secondly, below the band gap, the  $d_{yz}$ orbital has the same occupation  on  both sublattices, while the  $d_{xy}$ orbital is occupied on one sublattice and the  $d_{zx}$ orbital on the other.  
This pattern  in which  certain $d$ orbitals  are  preferentially occupied at different sites deviates from the case without orbital ordering,  in which  each of the  $d_{xy}$,  $d_{yz}$, and  $d_{zx}$  orbitals have the same occupancies on both Os sites, as shown in Ref. 27. These orbital occupations are consistent with mean field predictions of the occupancy of the Os $d$ orbitals at zero temperature, which also predict a staggered pattern  \cite{trivedi_2017}.  This staggered pattern arises from BNOO's distinctive blend of cFM order with strong SOC.

 To study  this ordering in greater depth, we can compute the occupation matrices, which  after diagonalization,  yield the occupation number (ON) eigenvalues and corresponding natural orbital (NO) eigenvectors. For a given Os atom, the $5d$ spin-orbitals have unequal amplitudes in each NO, as expected for the AFOO. The NOs also all have different occupation numbers. Regardless of their precise occupations, the unequal spin-orbital superpositions in the NOs endow the Os with a net non-zero spin and orbital moment. We moreover note that, due to \mbox{$5d$-$2p$} hybridization, there is significant charge transfer from O to Os, such that the charge on the $5d$ shell of Osmium is \mbox{$\langle n_d \rangle \approx 5$-$6$}, which is very different from the nominal heptavalent $5d^1$ filling from simple valence counting. 
Furthermore, the ten NOs are fractionally occupied with the largest ON close to $\langle n_1 \rangle \approx 1.0$ and the other nine NOs having occupations ranging from $0.37 - 0.56$. For the NO, $|1 \rangle$, with the occupation $\langle n_1 \rangle \approx 1.0$, the coefficients of the $e_g$ orbitals are an order of magnitude smaller than those for the $t_{2g}$ orbitals.

 \onecolumngrid
 \begin{center}	
 %
\begin{figure}[h]
  \vspace*{-0.3cm}
\begin{minipage}{0.98\hsize}
 \centerline{\includegraphics[scale=0.77]{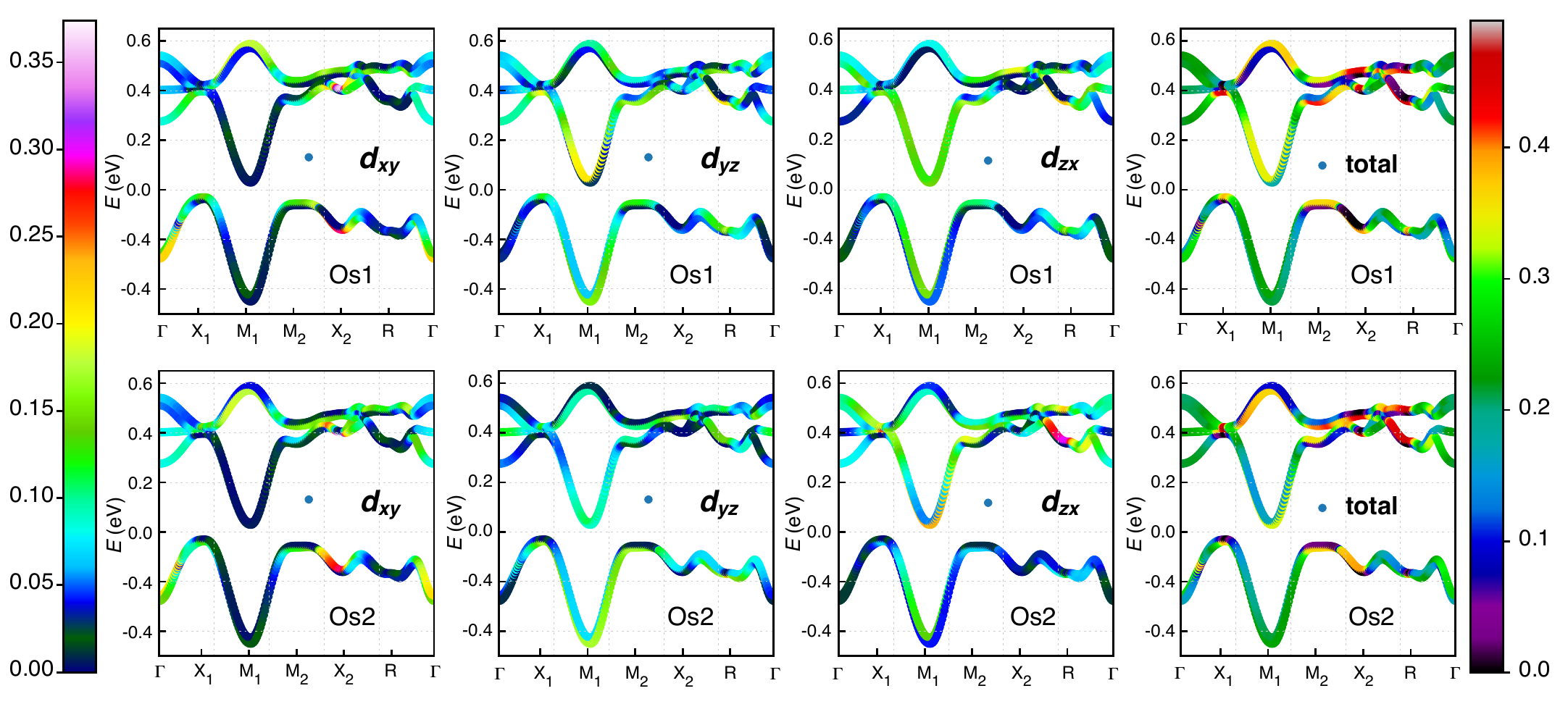}} 
\begin{minipage}{.98\hsize}
 \vspace*{-0.2cm}
\caption[]{\label{fig:bs} \small 
The band structures of the two sublattice Os atoms near the Fermi level with $5d$ partial characters for the  Q2 orthorhombic distortion (Model A.3 in Ref. \cite{CongEFG19}): Os1 sublattice  (top), Os2 sublattice   (bottom). The projection of each $5d$ orbital onto the Kohn-Sham bands is represented by the color shading. The color bar on the left shows the color scaling for the partial characters of the $t_{2g}$ orbitals, while the color bar on the right shows the scaling for all of the orbitals. The chosen high symmetry points are $\Gamma$=(0,0,0), X$_1$=($\frac{1}{2}$,0,0), M$_1$=(0,$\frac{1}{2}$,0), M$_2$=($\frac{1}{2}$,0,$\frac{1}{2})$, X$_2$=(0,0,$\frac{1}{2}$),  and R=(0,$\frac{1}{2}$,$\frac{1}{2}$). }
\end{minipage}
\end{minipage}
\end{figure}
%
\end{center}
   \vspace*{-0.80cm}
\twocolumngrid

In \mbox{Fig. \ref{fig:bs}}, we plot the band structures of the two sublattice Os atoms along the high symmetry directions of the monoclinic cell, with the total partial characters of the Os $5d$ bands color-coded proportional to their squared-amplitude contributions to the Kohn-Sham eigenvectors, the so-called fat-bands. The total partial character is the root of the sum of the squares of the partial characters of the Cartesian spin projections, $\sqrt{M_x^2+M_y^2+M_z^2}$. We plot the partial characters of the spin projections of $M_x$ and $M_y$ in the Supplemental Information Figs. 3 and 4, but not $M_z$ because it is two  orders of  magnitude smaller than the other two. For both Os atoms  along  these high symmetry directions, only $t_{2g}$ orbitals are occupied, consistent with $\langle n_d \rangle < 6$. The $t_{2g}$ and $e_g$ are irreducible representations of perfect cubic, octahedral, or tetrahedral symmetry. Because these symmetries are broken in the structure with Q2 distortion, there are no pure $t_{2g}$ or $e_g$ orbitals, nor a $\Delta(e_g - t_{2g})$ energy splitting, and there will be a small mixing between the two sets of orbitals.

We see that for both sublattices, below the band gap, the d$_{yz}$ orbital is  most heavily  occupied  (as denoted by the brighter green color), especially along the 
\mbox{X$_1$-M$_2$} direction, while  the  d$_{xy}$ and d$_{zx}$  orbitals  are less occupied. However, around the $\Gamma$ point, the $d_{xy}$ orbital  obviously has the largest occupancy.  We point out that here  the $d$ orbital character contribution  is only for the selected high symmetry directions. Thus,  it can not be directly compared with the PDOS result. Nevertheless, the different orbital character contributions reflected in the color can also be observed for all three $t_{2g}$ orbitals, especially  along the X$_2$-$\Gamma$ line. We can also see that the dispersions are largest along the X$_1$-M$_2$ and X$_1$-$\Gamma$ paths, while the bands are  flatter  from M$_2$ to R. 
The band gap is indirect and $\approx 0.06$ eV in magnitude. 

Finally, we have computed the gaps for the imposed cFM phase. We found that the gaps for the cFM phase with the DFT+U parameters of $U = 3.3$ eV and $J = 0.5$ eV are finite, but too small to be considered Mott insulating gaps. However, we find that the gap opens dramatically as we raise $U$ to 5.0 eV, as  shown in  detail in the  Supplemental Information.   In fact, even a ``small" increase to $U  = 4.0 \,  {\rm eV}$  is sufficient to open the gap to $E_{\rm gap} = 0.244 \,  {\rm eV}$. This indicates that the true value of $U$ for the osmium  $5d$ shell in BNOO could plausibly approach $4.0$ eV, but not exceed it. Previously, it was found that  LDA+U with $U = 4\, {\rm eV} > W$ was insufficient to  open a gap \cite{Pickett_2007}. Here, we demonstrated that GGA+SOC+U is sufficient to open a gap for $U\approx 4.0 \, {\rm eV}$.

 \section{Conclusions}
\label{Conc}
  \vspace*{-0.20cm}
In this work, we carried out DFT+U calculations on the magnetic Mott insulator Ba$_2$NaOsO$_6$, which has strong spin orbit coupling. Our numerical work is inspired by our recent NMR results revealing that this material exhibited a broken local point symmetry  (BLPS) phase followed by a two-sublattice  exotic canted ferromagnetic order (cFM). The local symmetry is broken by the orthorhombic Q2 distortion mode \cite{CongEFG19}. The question  we addressed  here  is whether this distortion  
 is accompanied by the  emergence of orbital order.  It was previously   proposed that the two-sublattice magnetic structure, revealed by NMR,  is the  very manifestation of staggered quadrupolar order with distinct orbital polarization on  the two sublattices arising from multipolar  exchange interactions \cite{Chen_PRB_2010, Lu_NatureComm_2017}. Moreover, it was indicated via a different mean field formalism that the anisotropic interactions result in orbital order that stabilizes exotic magnetic order   \cite{trivedi_2017}. Therefore, distinct  mean field  approaches  \cite{Chen_PRB_2010, trivedi_2017}  with a common ingredient of anisotropic exchange interactions imply that exotic magnetic order, such as the cFM reported in Ref. 12,  is  accompanied/driven by an orbital order. 

Motivated by the cFM order detected in NMR experiments, here we   investigated BNOO's orbital ordering pattern from  first  principles. We found  two-sublattice  orbital ordering,  illustrated by the spin density plots, within the alternating planes in which the total magnetic moment resides.  An auxiliary signature of the orbital ordering is revealed  by  the occupancies of the $t_{2g}$ orbitals in the  density  of states and  band structures. Our first principles work demonstrates that this  two-sublattice orbital ordering mainly arises from cFM order and strong SOC.
Moving forward, it would be worthwhile to more thoroughly investigate the cFM order observed in this work using other functionals or methods more adept at handling strong correlation to eliminate any ambiguities that stem from our specific computational treatment.

  \section{Acknowledgments}
\label{Ack}

The authors thank Jeong-Pil Song and Yiou Zhang  for enlightening discussions. We are especially grateful to Ian Fisher for  his  long term collaboration on the physics of \BaOs. This work was supported in part by U.S. National Science Foundation grants DMR-1608760 (V.F.M.) and DMR-1726213 (B.M.R.). The calculations presented here were performed using resources at the Center for Computation and Visualization, Brown University, which is supported by NSF Grant No. ACI-1548562.

 $^\dag$ Corresponding authors V. F. M. (vemi@brown.edu)  \&  B. R. (brenda\underline{ }rubenstein@brown.edu)

\bibliography{ref_OOEvid}

\begin{thebibliography}{27}%
\makeatletter
\providecommand \@ifxundefined [1]{%
 \@ifx{#1\undefined}
}%
\providecommand \@ifnum [1]{%
 \ifnum #1\expandafter \@firstoftwo
 \else \expandafter \@secondoftwo
 \fi
}%
\providecommand \@ifx [1]{%
 \ifx #1\expandafter \@firstoftwo
 \else \expandafter \@secondoftwo
 \fi
}%
\providecommand \natexlab [1]{#1}%
\providecommand \enquote  [1]{``#1''}%
\providecommand \bibnamefont  [1]{#1}%
\providecommand \bibfnamefont [1]{#1}%
\providecommand \citenamefont [1]{#1}%
\providecommand \href@noop [0]{\@secondoftwo}%
\providecommand \href [0]{\begingroup \@sanitize@url \@href}%
\providecommand \@href[1]{\@@startlink{#1}\@@href}%
\providecommand \@@href[1]{\endgroup#1\@@endlink}%
\providecommand \@sanitize@url [0]{\catcode `\\12\catcode `\$12\catcode
  `\&12\catcode `\#12\catcode `\^12\catcode `\_12\catcode `\%12\relax}%
\providecommand \@@startlink[1]{}%
\providecommand \@@endlink[0]{}%
\providecommand \url  [0]{\begingroup\@sanitize@url \@url }%
\providecommand \@url [1]{\endgroup\@href {#1}{\urlprefix }}%
\providecommand \urlprefix  [0]{URL }%
\providecommand \Eprint [0]{\href }%
\providecommand \doibase [0]{http://dx.doi.org/}%
\providecommand \selectlanguage [0]{\@gobble}%
\providecommand \bibinfo  [0]{\@secondoftwo}%
\providecommand \bibfield  [0]{\@secondoftwo}%
\providecommand \translation [1]{[#1]}%
\providecommand \BibitemOpen [0]{}%
\providecommand \bibitemStop [0]{}%
\providecommand \bibitemNoStop [0]{.\EOS\space}%
\providecommand \EOS [0]{\spacefactor3000\relax}%
\providecommand \BibitemShut  [1]{\csname bibitem#1\endcsname}%
\let\auto@bib@innerbib\@empty
\bibitem [{\citenamefont {Kim}\ \emph {et~al.}(2008)\citenamefont {Kim},
  \citenamefont {Jin}, \citenamefont {Moon}, \citenamefont {Kim}, \citenamefont
  {Park}, \citenamefont {Leem}, \citenamefont {Yu}, \citenamefont {Noh},
  \citenamefont {Kim}, \citenamefont {Oh}, \citenamefont {Park}, \citenamefont
  {Durairaj}, \citenamefont {Cao},\ and\ \citenamefont {Rotenberg}}]{Kim08Nov}%
  \BibitemOpen
  \bibfield  {author} {\bibinfo {author} {\bibfnamefont {B.~J.}\ \bibnamefont
  {Kim}}, \bibinfo {author} {\bibfnamefont {H.}~\bibnamefont {Jin}}, \bibinfo
  {author} {\bibfnamefont {S.~J.}\ \bibnamefont {Moon}}, \bibinfo {author}
  {\bibfnamefont {J.-Y.}\ \bibnamefont {Kim}}, \bibinfo {author} {\bibfnamefont
  {B.-G.}\ \bibnamefont {Park}}, \bibinfo {author} {\bibfnamefont {C.~S.}\
  \bibnamefont {Leem}}, \bibinfo {author} {\bibfnamefont {J.}~\bibnamefont
  {Yu}}, \bibinfo {author} {\bibfnamefont {T.~W.}\ \bibnamefont {Noh}},
  \bibinfo {author} {\bibfnamefont {C.}~\bibnamefont {Kim}}, \bibinfo {author}
  {\bibfnamefont {S.-J.}\ \bibnamefont {Oh}}, \bibinfo {author} {\bibfnamefont
  {J.-H.}\ \bibnamefont {Park}}, \bibinfo {author} {\bibfnamefont
  {V.}~\bibnamefont {Durairaj}}, \bibinfo {author} {\bibfnamefont
  {G.}~\bibnamefont {Cao}}, \ and\ \bibinfo {author} {\bibfnamefont
  {E.}~\bibnamefont {Rotenberg}},\ }\href@noop {} {\bibfield  {journal}
  {\bibinfo  {journal} {Phys. Rev. Lett.}\ }\textbf {\bibinfo {volume} {101}},\
  \bibinfo {pages} {076402} (\bibinfo {year} {2008})}\BibitemShut {NoStop}%
\bibitem [{\citenamefont {Chen}\ \emph {et~al.}(2010)\citenamefont {Chen},
  \citenamefont {Pereira},\ and\ \citenamefont {Balents}}]{Chen_PRB_2010}%
  \BibitemOpen
  \bibfield  {author} {\bibinfo {author} {\bibfnamefont {G.}~\bibnamefont
  {Chen}}, \bibinfo {author} {\bibfnamefont {R.}~\bibnamefont {Pereira}}, \
  and\ \bibinfo {author} {\bibfnamefont {L.}~\bibnamefont {Balents}},\
  }\href@noop {} {\bibfield  {journal} {\bibinfo  {journal} {Phys. Rev. B}\
  }\textbf {\bibinfo {volume} {82}},\ \bibinfo {pages} {174440} (\bibinfo
  {year} {2010})}\BibitemShut {NoStop}%
\bibitem [{\citenamefont {Chen}\ and\ \citenamefont
  {Balents}(2011)}]{Chen:2011}%
  \BibitemOpen
  \bibfield  {author} {\bibinfo {author} {\bibfnamefont {G.}~\bibnamefont
  {Chen}}\ and\ \bibinfo {author} {\bibfnamefont {L.}~\bibnamefont {Balents}},\
  }\href@noop {} {\bibfield  {journal} {\bibinfo  {journal} {Physical Review
  B}\ }\textbf {\bibinfo {volume} {84}},\ \bibinfo {pages} {094420} (\bibinfo
  {year} {2011})}\BibitemShut {NoStop}%
\bibitem [{\citenamefont {Ishizuka}\ and\ \citenamefont
  {Balents}(2014)}]{Ishizuka:2014el}%
  \BibitemOpen
  \bibfield  {author} {\bibinfo {author} {\bibfnamefont {H.}~\bibnamefont
  {Ishizuka}}\ and\ \bibinfo {author} {\bibfnamefont {L.}~\bibnamefont
  {Balents}},\ }\href@noop {} {\bibfield  {journal} {\bibinfo  {journal}
  {Physical Review B}\ }\textbf {\bibinfo {volume} {90}},\ \bibinfo {pages}
  {184422} (\bibinfo {year} {2014})}\BibitemShut {NoStop}%
\bibitem [{\citenamefont {Svoboda}\ \emph {et~al.}(2017)\citenamefont
  {Svoboda}, \citenamefont {Randeria},\ and\ \citenamefont
  {Trivedi}}]{trivedi_2017}%
  \BibitemOpen
  \bibfield  {author} {\bibinfo {author} {\bibfnamefont {C.}~\bibnamefont
  {Svoboda}}, \bibinfo {author} {\bibfnamefont {M.}~\bibnamefont {Randeria}}, \
  and\ \bibinfo {author} {\bibfnamefont {N.}~\bibnamefont {Trivedi}},\
  }\href@noop {} {\bibfield  {journal} {\bibinfo  {journal} {arXiv:1702.03199v1
  (unpublished)}\ } (\bibinfo {year} {2017})}\BibitemShut {NoStop}%
\bibitem [{\citenamefont {Witczak-Krempa}\ \emph {et~al.}(2014)\citenamefont
  {Witczak-Krempa}, \citenamefont {Chen}, \citenamefont {Kim},\ and\
  \citenamefont {Balents}}]{balents_SOC_review_2014}%
  \BibitemOpen
  \bibfield  {author} {\bibinfo {author} {\bibfnamefont {W.}~\bibnamefont
  {Witczak-Krempa}}, \bibinfo {author} {\bibfnamefont {G.}~\bibnamefont
  {Chen}}, \bibinfo {author} {\bibfnamefont {Y.~B.}\ \bibnamefont {Kim}}, \
  and\ \bibinfo {author} {\bibfnamefont {L.}~\bibnamefont {Balents}},\
  }\href@noop {} {\bibfield  {journal} {\bibinfo  {journal} {Annual Review of
  Condensed Matter Physics}\ }\textbf {\bibinfo {volume} {5}},\ \bibinfo
  {pages} {57} (\bibinfo {year} {2014})}\BibitemShut {NoStop}%
\bibitem [{\citenamefont {Romh\'anyi}\ \emph {et~al.}(2017)\citenamefont
  {Romh\'anyi}, \citenamefont {Balents},\ and\ \citenamefont
  {Jackeli}}]{Balents_2017}%
  \BibitemOpen
  \bibfield  {author} {\bibinfo {author} {\bibfnamefont {J.}~\bibnamefont
  {Romh\'anyi}}, \bibinfo {author} {\bibfnamefont {L.}~\bibnamefont {Balents}},
  \ and\ \bibinfo {author} {\bibfnamefont {G.}~\bibnamefont {Jackeli}},\
  }\href@noop {} {\bibfield  {journal} {\bibinfo  {journal} {Phys. Rev. Lett.}\
  }\textbf {\bibinfo {volume} {118}},\ \bibinfo {pages} {217202} (\bibinfo
  {year} {2017})}\BibitemShut {NoStop}%
\bibitem [{\citenamefont {Kim}\ \emph {et~al.}(2009)\citenamefont {Kim},
  \citenamefont {Ohsumi}, \citenamefont {Komesu}, \citenamefont {Sakai},
  \citenamefont {Morita}, \citenamefont {Takagi},\ and\ \citenamefont
  {Arima}}]{J_eff_half_MI_2}%
  \BibitemOpen
  \bibfield  {author} {\bibinfo {author} {\bibfnamefont {B.~J.}\ \bibnamefont
  {Kim}}, \bibinfo {author} {\bibfnamefont {H.}~\bibnamefont {Ohsumi}},
  \bibinfo {author} {\bibfnamefont {T.}~\bibnamefont {Komesu}}, \bibinfo
  {author} {\bibfnamefont {S.}~\bibnamefont {Sakai}}, \bibinfo {author}
  {\bibfnamefont {T.}~\bibnamefont {Morita}}, \bibinfo {author} {\bibfnamefont
  {H.}~\bibnamefont {Takagi}}, \ and\ \bibinfo {author} {\bibfnamefont
  {T.}~\bibnamefont {Arima}},\ }\href@noop {} {\bibfield  {journal} {\bibinfo
  {journal} {Science}\ }\textbf {\bibinfo {volume} {323}},\ \bibinfo {pages}
  {1329} (\bibinfo {year} {2009})}\BibitemShut {NoStop}%
\bibitem [{\citenamefont {Zhang}\ \emph {et~al.}(2013)\citenamefont {Zhang},
  \citenamefont {Haule},\ and\ \citenamefont
  {Vanderbilt}}]{Jeff_half_iridates}%
  \BibitemOpen
  \bibfield  {author} {\bibinfo {author} {\bibfnamefont {H.}~\bibnamefont
  {Zhang}}, \bibinfo {author} {\bibfnamefont {K.}~\bibnamefont {Haule}}, \ and\
  \bibinfo {author} {\bibfnamefont {D.}~\bibnamefont {Vanderbilt}},\
  }\href@noop {} {\bibfield  {journal} {\bibinfo  {journal} {Phys. Rev. Lett.}\
  }\textbf {\bibinfo {volume} {111}},\ \bibinfo {pages} {246402} (\bibinfo
  {year} {2013})}\BibitemShut {NoStop}%
\bibitem [{\citenamefont {Moon}\ \emph {et~al.}(2008)\citenamefont {Moon},
  \citenamefont {Jin}, \citenamefont {Kim}, \citenamefont {Choi}, \citenamefont
  {Lee}, \citenamefont {Yu}, \citenamefont {Cao}, \citenamefont {Sumi},
  \citenamefont {Funakubo}, \citenamefont {Bernhard},\ and\ \citenamefont
  {Noh}}]{5d_IMT_CMS_RP_series}%
  \BibitemOpen
  \bibfield  {author} {\bibinfo {author} {\bibfnamefont {S.~J.}\ \bibnamefont
  {Moon}}, \bibinfo {author} {\bibfnamefont {H.}~\bibnamefont {Jin}}, \bibinfo
  {author} {\bibfnamefont {K.~W.}\ \bibnamefont {Kim}}, \bibinfo {author}
  {\bibfnamefont {W.~S.}\ \bibnamefont {Choi}}, \bibinfo {author}
  {\bibfnamefont {Y.~S.}\ \bibnamefont {Lee}}, \bibinfo {author} {\bibfnamefont
  {J.}~\bibnamefont {Yu}}, \bibinfo {author} {\bibfnamefont {G.}~\bibnamefont
  {Cao}}, \bibinfo {author} {\bibfnamefont {A.}~\bibnamefont {Sumi}}, \bibinfo
  {author} {\bibfnamefont {H.}~\bibnamefont {Funakubo}}, \bibinfo {author}
  {\bibfnamefont {C.}~\bibnamefont {Bernhard}}, \ and\ \bibinfo {author}
  {\bibfnamefont {T.~W.}\ \bibnamefont {Noh}},\ }\href@noop {} {\bibfield
  {journal} {\bibinfo  {journal} {Phys. Rev. Lett.}\ }\textbf {\bibinfo
  {volume} {101}},\ \bibinfo {pages} {226402} (\bibinfo {year}
  {2008})}\BibitemShut {NoStop}%
\bibitem [{\citenamefont {Wan}\ \emph {et~al.}(2011)\citenamefont {Wan},
  \citenamefont {Turner}, \citenamefont {Vishwanath},\ and\ \citenamefont
  {Savrasov}}]{weyl}%
  \BibitemOpen
  \bibfield  {author} {\bibinfo {author} {\bibfnamefont {X.}~\bibnamefont
  {Wan}}, \bibinfo {author} {\bibfnamefont {A.~M.}\ \bibnamefont {Turner}},
  \bibinfo {author} {\bibfnamefont {A.}~\bibnamefont {Vishwanath}}, \ and\
  \bibinfo {author} {\bibfnamefont {S.~Y.}\ \bibnamefont {Savrasov}},\
  }\href@noop {} {\bibfield  {journal} {\bibinfo  {journal} {Phys. Rev. B}\
  }\textbf {\bibinfo {volume} {83}},\ \bibinfo {pages} {205101} (\bibinfo
  {year} {2011})}\BibitemShut {NoStop}%
\bibitem [{\citenamefont {Lu}\ \emph {et~al.}(2017)\citenamefont {Lu},
  \citenamefont {Song}, \citenamefont {Liu}, \citenamefont {Reyes},
  \citenamefont {Kuhns}, \citenamefont {Lee}, \citenamefont {Fisher},\ and\
  \citenamefont {Mitrovi{\'c}}}]{Lu_NatureComm_2017}%
  \BibitemOpen
  \bibfield  {author} {\bibinfo {author} {\bibfnamefont {L.}~\bibnamefont
  {Lu}}, \bibinfo {author} {\bibfnamefont {M.}~\bibnamefont {Song}}, \bibinfo
  {author} {\bibfnamefont {W.}~\bibnamefont {Liu}}, \bibinfo {author}
  {\bibfnamefont {A.~P.}\ \bibnamefont {Reyes}}, \bibinfo {author}
  {\bibfnamefont {P.}~\bibnamefont {Kuhns}}, \bibinfo {author} {\bibfnamefont
  {H.~O.}\ \bibnamefont {Lee}}, \bibinfo {author} {\bibfnamefont {I.~R.}\
  \bibnamefont {Fisher}}, \ and\ \bibinfo {author} {\bibfnamefont {V.~F.}\
  \bibnamefont {Mitrovi{\'c}}},\ }\href@noop {} {\bibfield  {journal} {\bibinfo
   {journal} {Nature Communications}\ }\textbf {\bibinfo {volume} {8}},\
  \bibinfo {pages} {14407 EP } (\bibinfo {year} {2017})}\BibitemShut {NoStop}%
\bibitem [{\citenamefont {Kresse}\ and\ \citenamefont {Hafner}(1993)}]{vasp_1}%
  \BibitemOpen
  \bibfield  {author} {\bibinfo {author} {\bibfnamefont {G.}~\bibnamefont
  {Kresse}}\ and\ \bibinfo {author} {\bibfnamefont {J.}~\bibnamefont
  {Hafner}},\ }\href@noop {} {\bibfield  {journal} {\bibinfo  {journal} {Phys.
  Rev. B}\ }\textbf {\bibinfo {volume} {47}},\ \bibinfo {pages} {558} (\bibinfo
  {year} {1993})}\BibitemShut {NoStop}%
\bibitem [{\citenamefont {Kresse}\ and\ \citenamefont {Hafner}(1994)}]{vasp_2}%
  \BibitemOpen
  \bibfield  {author} {\bibinfo {author} {\bibfnamefont {G.}~\bibnamefont
  {Kresse}}\ and\ \bibinfo {author} {\bibfnamefont {J.}~\bibnamefont
  {Hafner}},\ }\href@noop {} {\bibfield  {journal} {\bibinfo  {journal} {Phys.
  Rev. B}\ }\textbf {\bibinfo {volume} {49}},\ \bibinfo {pages} {251} (\bibinfo
  {year} {1994})}\BibitemShut {NoStop}%
\bibitem [{\citenamefont {Kresse}\ and\ \citenamefont
  {Furthmuller}(1996{\natexlab{a}})}]{vasp_3}%
  \BibitemOpen
  \bibfield  {author} {\bibinfo {author} {\bibfnamefont {G.}~\bibnamefont
  {Kresse}}\ and\ \bibinfo {author} {\bibfnamefont {J.}~\bibnamefont
  {Furthmuller}},\ }\href@noop {} {\bibfield  {journal} {\bibinfo  {journal}
  {Comput. Mat. Sci.}\ }\textbf {\bibinfo {volume} {6}},\ \bibinfo {pages} {15}
  (\bibinfo {year} {1996}{\natexlab{a}})}\BibitemShut {NoStop}%
\bibitem [{\citenamefont {Kresse}\ and\ \citenamefont
  {Furthmuller}(1996{\natexlab{b}})}]{vasp_4}%
  \BibitemOpen
  \bibfield  {author} {\bibinfo {author} {\bibfnamefont {G.}~\bibnamefont
  {Kresse}}\ and\ \bibinfo {author} {\bibfnamefont {J.}~\bibnamefont
  {Furthmuller}},\ }\href@noop {} {\bibfield  {journal} {\bibinfo  {journal}
  {Phys. Rev. B}\ }\textbf {\bibinfo {volume} {54}},\ \bibinfo {pages} {11169}
  (\bibinfo {year} {1996}{\natexlab{b}})}\BibitemShut {NoStop}%
\bibitem [{\citenamefont {Perdew}\ and\ \citenamefont {Wang}(1992)}]{GGA}%
  \BibitemOpen
  \bibfield  {author} {\bibinfo {author} {\bibfnamefont {J.~P.}\ \bibnamefont
  {Perdew}}\ and\ \bibinfo {author} {\bibfnamefont {Y.}~\bibnamefont {Wang}},\
  }\href@noop {} {\bibfield  {journal} {\bibinfo  {journal} {Phys. Rev. B}\
  }\textbf {\bibinfo {volume} {45}},\ \bibinfo {pages} {13244} (\bibinfo {year}
  {1992})}\BibitemShut {NoStop}%
\bibitem [{\citenamefont {Hund}()}]{hund}%
  \BibitemOpen
  \bibfield  {author} {\bibinfo {author} {\bibfnamefont {F.~Z.}\ \bibnamefont
  {Hund}},\ }\href@noop {} {\bibfield  {journal} {\bibinfo  {journal}
  {Zeitschrift f{\"u}r Physik}\ }\textbf {\bibinfo {volume} {33}},\ \bibinfo
  {pages} {345}}\BibitemShut {NoStop}%
\bibitem [{\citenamefont {Bl{\"o}chl}(1994)}]{PAW_Blochl}%
  \BibitemOpen
  \bibfield  {author} {\bibinfo {author} {\bibfnamefont {P.}~\bibnamefont
  {Bl{\"o}chl}},\ }\href@noop {} {\bibfield  {journal} {\bibinfo  {journal}
  {Phys. Rev. B}\ }\textbf {\bibinfo {volume} {50}},\ \bibinfo {pages} {17953}
  (\bibinfo {year} {1994})}\BibitemShut {NoStop}%
\bibitem [{\citenamefont {Kresse}\ and\ \citenamefont
  {Joubert}(1999)}]{PAW_vasp}%
  \BibitemOpen
  \bibfield  {author} {\bibinfo {author} {\bibfnamefont {G.}~\bibnamefont
  {Kresse}}\ and\ \bibinfo {author} {\bibfnamefont {J.}~\bibnamefont
  {Joubert}},\ }\href@noop {} {\bibfield  {journal} {\bibinfo  {journal} {Phys.
  Rev. B}\ }\textbf {\bibinfo {volume} {59}},\ \bibinfo {pages} {1758}
  (\bibinfo {year} {1999})}\BibitemShut {NoStop}%
\bibitem [{\citenamefont {{Cong}}\ \emph {et~al.}(2019)\citenamefont {{Cong}},
  \citenamefont {{Nanguneri}}, \citenamefont {{Rubenstein}},\ and\
  \citenamefont {{Mitrovic}}}]{CongEFG19}%
  \BibitemOpen
  \bibfield  {author} {\bibinfo {author} {\bibfnamefont {R.}~\bibnamefont
  {{Cong}}}, \bibinfo {author} {\bibfnamefont {R.}~\bibnamefont {{Nanguneri}}},
  \bibinfo {author} {\bibfnamefont {B.~M.}\ \bibnamefont {{Rubenstein}}}, \
  and\ \bibinfo {author} {\bibfnamefont {V.~F.}\ \bibnamefont {{Mitrovic}}},\
  }\href@noop {} {\bibfield  {journal} {\bibinfo  {journal} {arXiv e-prints}\
  ,\ \bibinfo {eid} {arXiv:1908.09014}} (\bibinfo {year} {2019})},\ \Eprint
  {http://arxiv.org/abs/1908.09014} {arXiv:1908.09014} \BibitemShut {NoStop}%
\bibitem [{\citenamefont {Erickson}\ \emph {et~al.}(2007)\citenamefont
  {Erickson}, \citenamefont {Misra}, \citenamefont {Miller}, \citenamefont
  {Gupta}, \citenamefont {Schlesinger}, \citenamefont {Harrison}, \citenamefont
  {Kim},\ and\ \citenamefont {Fisher}}]{erickson2007FM}%
  \BibitemOpen
  \bibfield  {author} {\bibinfo {author} {\bibfnamefont {A.}~\bibnamefont
  {Erickson}}, \bibinfo {author} {\bibfnamefont {S.}~\bibnamefont {Misra}},
  \bibinfo {author} {\bibfnamefont {G.~J.}\ \bibnamefont {Miller}}, \bibinfo
  {author} {\bibfnamefont {R.}~\bibnamefont {Gupta}}, \bibinfo {author}
  {\bibfnamefont {Z.}~\bibnamefont {Schlesinger}}, \bibinfo {author}
  {\bibfnamefont {W.}~\bibnamefont {Harrison}}, \bibinfo {author}
  {\bibfnamefont {J.}~\bibnamefont {Kim}}, \ and\ \bibinfo {author}
  {\bibfnamefont {I.}~\bibnamefont {Fisher}},\ }\href@noop {} {\bibfield
  {journal} {\bibinfo  {journal} {Physical Review Letters}\ }\textbf {\bibinfo
  {volume} {99}},\ \bibinfo {pages} {016404} (\bibinfo {year}
  {2007})}\BibitemShut {NoStop}%
\bibitem [{\citenamefont {Methfessel}\ and\ \citenamefont
  {Paxton}(1989)}]{MP_smearing}%
  \BibitemOpen
  \bibfield  {author} {\bibinfo {author} {\bibfnamefont {M.}~\bibnamefont
  {Methfessel}}\ and\ \bibinfo {author} {\bibfnamefont {A.~T.}\ \bibnamefont
  {Paxton}},\ }\href@noop {} {\bibfield  {journal} {\bibinfo  {journal} {Phys.
  Rev. B}\ }\textbf {\bibinfo {volume} {40}},\ \bibinfo {pages} {3616}
  (\bibinfo {year} {1989})}\BibitemShut {NoStop}%
\bibitem [{\citenamefont {Bl\"ochl}\ \emph {et~al.}(1994)\citenamefont
  {Bl\"ochl}, \citenamefont {Jepsen},\ and\ \citenamefont
  {Andersen}}]{Blochl_tetrahedron_corrections}%
  \BibitemOpen
  \bibfield  {author} {\bibinfo {author} {\bibfnamefont {P.~E.}\ \bibnamefont
  {Bl\"ochl}}, \bibinfo {author} {\bibfnamefont {O.}~\bibnamefont {Jepsen}}, \
  and\ \bibinfo {author} {\bibfnamefont {O.~K.}\ \bibnamefont {Andersen}},\
  }\href@noop {} {\bibfield  {journal} {\bibinfo  {journal} {Phys. Rev. B}\
  }\textbf {\bibinfo {volume} {49}},\ \bibinfo {pages} {16223} (\bibinfo {year}
  {1994})}\BibitemShut {NoStop}%
\bibitem [{\citenamefont {Mosca}(2019)}]{thesisCF}%
  \BibitemOpen
  \bibfield  {author} {\bibinfo {author} {\bibfnamefont {D.~F.}\ \bibnamefont
  {Mosca}},\ }\href@noop {} {\enquote {\bibinfo {title} {{Master Thesis -
  Universit{\`a} di Bologna: Quantum Magnetism in Relativistic Osmates from
  First Principles}},}\ } (\bibinfo {year} {2019})\BibitemShut {NoStop}%
\bibitem [{\citenamefont {Gangopadhyay}\ and\ \citenamefont
  {Pickett}(2015)}]{Pickett_2015}%
  \BibitemOpen
  \bibfield  {author} {\bibinfo {author} {\bibfnamefont {S.}~\bibnamefont
  {Gangopadhyay}}\ and\ \bibinfo {author} {\bibfnamefont {W.~E.}\ \bibnamefont
  {Pickett}},\ }\href@noop {} {\bibfield  {journal} {\bibinfo  {journal} {Phys.
  Rev. B}\ }\textbf {\bibinfo {volume} {91}},\ \bibinfo {pages} {045133}
  (\bibinfo {year} {2015})}\BibitemShut {NoStop}%
\bibitem [{\citenamefont {Lee}\ and\ \citenamefont
  {Pickett}(2007)}]{Pickett_2007}%
  \BibitemOpen
  \bibfield  {author} {\bibinfo {author} {\bibfnamefont {K.-W.}\ \bibnamefont
  {Lee}}\ and\ \bibinfo {author} {\bibfnamefont {W.~E.}\ \bibnamefont
  {Pickett}},\ }\href@noop {} {\bibfield  {journal} {\bibinfo  {journal} {EPL
  (Europhysics Letters)}\ }\textbf {\bibinfo {volume} {80}},\ \bibinfo {pages}
  {37008} (\bibinfo {year} {2007})}\BibitemShut {NoStop}%
\end{thebibliography}%

\end{document}